\documentclass[
amsmath,amssymb,aps,showpacs,twocolumn,longbibliography,citeautoscript,superscriptaddress,footinbib,
]{revtex4-2}

\usepackage[dvips]{graphicx}
\usepackage{dcolumn}
\usepackage{bm}
\usepackage{placeins}
\usepackage{ifthen}
\usepackage{eurosym}
\usepackage{float}
\usepackage{longtable}
\usepackage{hyperref}
\usepackage{footnote}
\usepackage{rotating}
\usepackage{mathtools}
\usepackage{amsmath}
\usepackage{amssymb}
\usepackage[usenames,dvipsnames]{color}

\begin{document}

\preprint{APS/123-QED}

\title{Weak ferromagnetism linked to the high-temperature spiral phase of \texorpdfstring{YBaCuFeO$_5$}{YBaCuFeO5}}

\author{J. Lyu}
\email{jike.lyu@psi.ch}
\affiliation{Laboratory for Multiscale Materials Experiments, Paul Scherrer Institut, 5232 Villigen PSI, Switzerland}
\author{M. Morin}
\affiliation{Laboratory for Multiscale Materials Experiments, Paul Scherrer Institut, 5232 Villigen PSI, Switzerland}
\affiliation{Excelsus Structural Solutions (Swiss) AG, PARK innovAARE, 5234 Villigen, Switzerland}
\author{T. Shang}
\affiliation{Key Laboratory of Polar Materials and Devices (MOE), School of Physics and Electronic Science, East China Normal University, Shanghai 200241, China}
\affiliation{Laboratory for Multiscale Materials Experiments, Paul Scherrer Institut, 5232 Villigen PSI, Switzerland}
\author{M. T. Fern\'{a}ndez-D\'{\i}az}
\affiliation{Institut Laue Langevin, 71 avenue des Martyrs, CS 20156, 38042 Grenoble Cedex 9, France}
\author{M. Medarde}
\email{marisa.medarde@psi.ch}
\affiliation{Laboratory for Multiscale Materials Experiments, Paul Scherrer Institut, 5232 Villigen PSI, Switzerland}

\date{\today}

\begin{abstract}
The layered perovskite YBaCuFeO$_5$ is a rare example of cycloidal spiral magnet whose ordering temperature T$_{spiral}$ can be tuned far beyond room temperature by adjusting the degree of Cu$^{2+}$/Fe$^{3+}$ chemical disorder in the structure. This unusual property qualifies this material as one of the most promising spin-driven multiferroic candidates. However, very little is known about the response of the spiral to magnetic fields, crucial for magnetoelectric cross-control applications. Using bulk magnetization and neutron powder diffraction measurements under magnetic fields up to 9 T we report here the first temperature-magnetic field phase diagram of this material. Besides revealing a strong stability of the spiral state, our data uncover the presence of weak ferromagnetism coexisting with the spiral modulation. Since ferromagnets can be easily manipulated with magnetic fields, this observation opens new perspectives for the control of the spiral orientation, directly linked to the polarization direction, as well as for a possible future use of this material in technological applications.
\end{abstract}

\maketitle

\section{Introduction}

Multiferroic materials, in particular those where electric polarization  coexists with magnetic order, have attracted a great deal of interest in fundamental physics.~\cite{Cheong_2007,Kimura_2007,Khomskii_2009,Fiebig_2016,Spaldin_2019} In parallel, some of them are in the center of attention of applied sciences due to the strong coupling between their electric and magnetic orders, which could be exploited in low-power magnetoelectric applications.~\cite{Champel_2008,Fiebig_2016,Pugach_2017,Spaldin_2019,Ramesh_2019} Among these materials, frustrated magnets with ordered spiral phases are particularly promising because the non-collinear spiral magnetic order breaks space-inversion symmetry and can induce spontaneous polarization $P$ ~\cite{Kimura_2003,Kimura_2007} through inverse Dzyaloshinsky-Moriya~\cite{Sergienko_2006,Mochizuki_2010} or spin-current mechanisms.~\cite{Katsura_2005} The common origin of polarization and spiral order guarantees a substantial coupling, which is highly desirable for applications. However, the low ordering temperatures of most spiral magnets (typically $<$ 50 K) make their implementation in real-life devices unpractical. Another major dissadvantage is that spiral phases are in general antiferromagnetic, making quite difficult reading and controlling the spiral helicity, directly related with the polarization direction.  

To date, only a few materials featuring either spontaneous, or magnetic-field-induced magnetization coexisting with spiral magnetic order in the same phase have been reported. Examples of them are the conical magnets CoCr$_2$O$_4$~\cite{Yamasaki_2006}, ZnCr$_2$Se$_4$~\cite{Murakawa_2008}, Ba$_2$Mg$_2$Fe$_{12}$O$_{22}$~\cite{Kitagawa_2010}, and Mn$_2$GeO$_4$~\cite{White_2012}. This last material is particularly interesting because in contrast with the other examples, spontaneous magnetization and spin-driven polarization appear along the same direction, and modest magnetic fields can reverse simultaneously both ferroic variables while keeping the domain structure unchanged~\cite{Leo_2018}. Unfortunately, this happens at very low temperatures ($\sim$ 6 K). It will be thus desirable to identify other materials with coupled magnetization and spiral magnetic order coexisting at higher temperatures.

The layered perovskite YBaCuFeO$_5$ object of the present study is one of rare exceptions where spiral magnetic order can be stabilized at temperatures beyond RT.~\cite{Morin_2016,Shang_2018} But what makes this material truly exceptional compared with other high-temperature spiral magnets ~\cite{Damay_2021,Kimura_2008,Greenblatt_2011,Kitagawa_2010} is the extraordinary tunability of the spiral ordering temperature ($T_{spiral}$), which can be increased from 154 to 310 K upon enhancing the degree of Cu$^{2+}$/Fe$^{3+}$ chemical disorder in the structure. As shown in Fig. 1a, these two cations occupy the bipyramidal units in the tetragonal layered perovskite unit cell (space group $P4mm$)~\cite{ErRakho_1988, Morin_2015}. Contrarily to Ba$^{2+}$ and Y$^{3+}$, which order in layers perpendicular to the \textbf{c} crystal axis, Cu$^{2+}$ and Fe$^{3+}$ display a very particular type of correlated disorder characterized by the preferential occupation of the bipyramids by ferromagnetically-coupled Cu-Fe pairs. These pairs are disordered in the structure, with the degree of disorder - and hence $T_{spiral}$ - being strongly dependent on the preparation method.~\cite{Morin_2016}

The magnetic structure expected from both, the Goodenough-Kanamori-Anderson rules of superexchange and the \textit{ab-initio} calculations of the exchange constants~\cite{Morin_2015} is described by the magnetic propagation \textbf{k$_c$} = (1/2, 1/2, 1/2). This collinear arrangement, shown in Fig. 1a, is indeed observed experimentally, but at only high temperatures ($T_{spiral} < T < T_{collinear}$ ~\cite{Caignaert_1995, Morin_2015}). At lower temperatures ($T <T_{spiral}$) it is replaced by an inclined spiral described by the propagation vector \textbf{k$_i$} = (1/2, 1/2, 1/2$\pm q$) and schematically shown in Fig. 1a ~\cite{Morin_2015}, whose cycloidal component could induce ferroelectricity.~\cite{Katsura_2005,Sergienko_2006,Mostovoy_2006,Mochizuki_2010} The observation of incommensurate magnetic order beyond RT in a crystal structure without any apparent source of frustration was puzzling, and the gigantic, positive impact of the Fe/Cu disorder in the stability of the spiral seemed at odds with traditional frustration mechanisms.~\cite{Morin_2015} Interestingly, both observations were recently rationalized in terms of a novel, disorder-based frustration mechanism based in the gigantic impact of a few Fe-Fe "defects" occupying the bipyramidal units, which host majoritarily Cu-Fe pairs.~\cite{Scaramucci_2018,Scaramucci_2020} Since the coupling $J_{Fe-Fe}$ between the two Fe$^{3+}$ moments in one such defects is antiferromagnetic and about 100 times stronger that the weakly ferromagnetic $J_{Cu-Fe}$ coupling, a tiny defect concentration (2 to 5$\%$ of the bipyramids occupied by Fe-Fe pairs) is enough to destroy the collinear order and stabilize a spiral where both, $T_{spiral}$ and the magnetic modulation parameter $q$ are proportional to the defect concentration.~\cite{Scaramucci_2020}

According to neutron powder difraction investigations ~\cite{Morin_2015,Morin_2016, Shang_2018,Zhang_2021}, the inclined spiral phase of YBaCuFeO$_5$ is antiferromagnetic without a net spontaneous magnetization. However, given the (usually strong) response of frustrated magnets to external magnetic fields, low-field incommensurate conical phases as those reported for ZnCr$_2$Se$_4$~\cite{Murakawa_2008} and some hexaferrites~\cite{Kimura_2012} could exist in YBaCuFeO$_5$ as well. Since the stability of the spiral phase can be extended from RT up to $\sim$ 400 K by replacing 40$\%$ of Ba by Sr, and the cycloidal component can be tuned in different ways,~\cite{Shang_2018,Zhang_2021} the presence of a field-induced conical phase in YBaCuFeO$_5$ with net magnetization could result in a material with interesting potential for magnetoelectric applications.~\cite{Liang_2021} Motivated by this possibility we report here the first systematic investigation of the YBaCuFeO$_5$ magnetic phase diagram in a broad temperature (2 - 300 K) and magnetic field range (0 - 9 T). Using magnetometry and powder neutron diffraction under magnetic field we reveal the presence of three distinct magnetic phases. Moreover, we uncover the existence of weak ferromagnetism coexisting with the spiral modulation that can be switched using modest magnetic fields.

\section{Experimental details}

\subsection{Sample synthesis}
The procedure for obtaining YBaCuFeO$_5$ ceramics with adjustable T$_{spiral}$ temperatures has been reported previously ~\cite{Morin_2015, Morin_2016, Shang_2018}. The polycrystalline sample used in this study was prepared by solid state reaction. As precursors we used high-purity stoichiometric amounts of Y$_2$O$_3$ (99.999 \%, Aldrich), BaCO$_3$ (99.997 \%, Alfa Aesar), CuO (99.995 \%, Alfa Aesar) and Fe$_2$O$_3$ (99.998 \%, Alfa Aesar) that were weighted, thoroughly mixed and fired at 1150 $^\circ$C for 50 h under oxygen gas flow. The resulting black powder was grounded again, pelletized under 4 kbar in the form of a long rod ($D$ $\sim$ 5.6 mm, $L$ $\sim$ 5 cm), then sintered at 1150 $^\circ$C for 50 h in air, and finally cooled down to room temperature at a rate of 300 K/h. According to ref.~\cite{Morin_2016}, this cooling rate should result in a $T_{spiral}$ value close to $\sim$ 240 K, and a $T_{collinear}$ $\sim$ 440 K. The sample purity was checked by laboratory X-ray powder diffraction (Brucker D8 Advance, Cu K$_\alpha$), that revealed an excellent crystallinity and confirmed the absence of impurity phases within the detection limit of this technique ($< 1\%$).

\subsection{Magnetometry}
Magnetization $M$ and $ac$ susceptibility measurements $\chi_{ac}$ were performed in a Superconducting Quantum Interference Device Magnetometer (MPMS-XL, 7 T), and a Physical Properties Measuring System (PPMS, 9 T) from Quantum Design. A cylindrical YBaCuFeO$_5$ pellet ($L$ $\sim$ 3.8 mm, $D$ $\sim$ 5.6 mm, $m$ $\sim$ 520.7 mg) was cut from the same cylinder used for powder neutron diffraction measurements and mounted on a transparent drinking straw. The temperature dependence of the magnetization $M(T)$ was investigated between 5 K and 300 K under magnetic fields up to 7.0 T using zero-field cooled warming (ZFC), field cooled cooling (FCC) and field cooled warming (FCW) experimental protocols while slowly ramping the temperature at a cooling/heating rate of 1 K/min. The field dependence of the magnetization $M(H)$ was measured at selected temperatures between 20 K to 300 K and magnetic fields between 0 and 9 T after cooling the sample in zero field. After each field scan, the magnetic field was set to zero before heating to the next temperature. Additionally, full $M-H$ hysteresis cycles with fields between -7 T and + 7 T were conducted at selected temperatures $T$ $<$ $T_{spiral}$. The temperature dependence of the $ac$ susceptibility $\chi_{ac}(T)$ was measured with an $ac$ field of 15 Oe at several frequencies between 100 and 10000 Hz. The measurement were conducted by heating after cooling the sample to 5 K in zero field.

\begin{figure*}[!thb]
\includegraphics[keepaspectratio=true,width=180 mm]{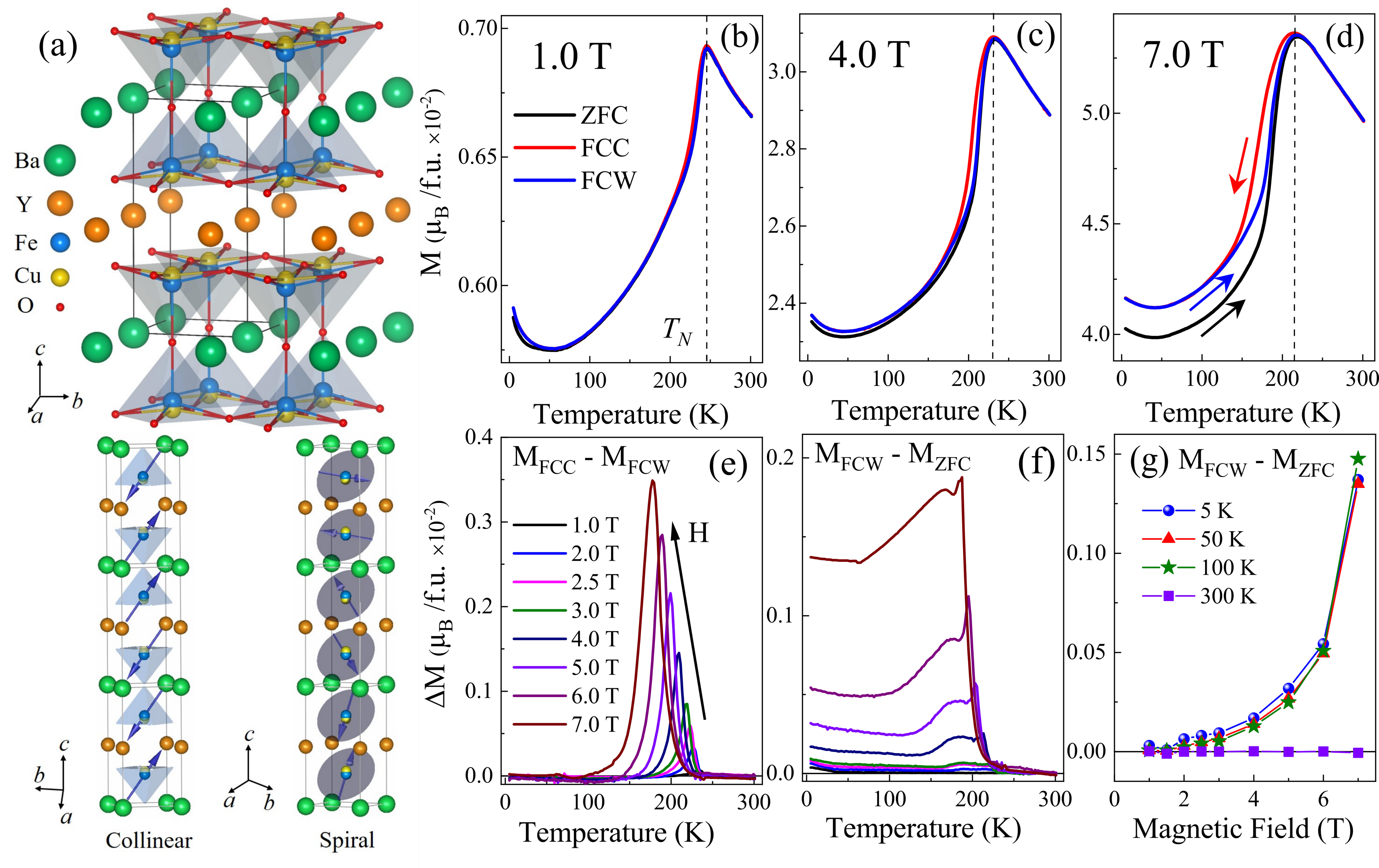}
\vspace{-3mm}
\caption{(a) Crystal and magnetic structures of YBaCuFeO$_5$. Upper panel: crystal structure illustrating the Cu/Fe chemical disorder within the bipyramidal units. The tetragonal $P4mm$ unit cell is indicated by the grey lines. Left down panel: high-temperature collinear magnetic structure with $\boldsymbol{k_c} = (1/2, 1/2, 1/2)$. Right down panel: low-temperature spiral magnetic structure with $\boldsymbol{k_i} = (1/2, 1/2, k_z)$. For sake of clarity reduced magnetic cells are shown in the last two panels. (b)--(d) Temperature dependence of the magnetization at 1, 4 and 7\,T measured under ZFC, FCC and FCW protocols. (e) Difference between the FCC and FCW magnetization curves $\Delta M_{hyst} = M_{FCC} - M_{FCW}$ showing the growth of the hysteretic region for increasing magnetic fields. (f) Difference between the FCW and ZFC magnetization curves $\Delta M_{irr} = M_{FCW} - M_{ZFC}$ showing the growth of the irreversible magnetization for increasing magnetic fields below $T_\mathrm{spiral}$. The color code is the same as in (e). (g) Magnetic field dependence of the irreversible magnetization $\Delta M_{irr} = M_{FCW} - M_{ZFC}$ at 5, 50, 100 and 300\,K.}
\label{Fig2:MvsT_pellet}
\end{figure*}

\subsection{Neutron Diffraction under magnetic field}
Powder neutron diffraction (PND) measurements were carried out at the Institut Laue Langevin (ILL) in Grenoble, France. To avoid the displacement of the crystallites under the action of the magnetic field we used a sintered YBaCuFeO$_5$ ceramic bar (see Sample Synthesis), that was fixed with aluminum foil inside of a cylindrical vanadium can. The can was mounted on the stick of a cryomagnet, and cooled down to 1.5 K. PND patterns were continuously recorded at the powder diffractometer D20 [Ge(224), take-off angle 90$^{\circ}$, 2$\theta_{max}$ = 140$^{\circ}$, 2$\theta_{step}$ = 0.05$^{\circ}$, $\lambda$ = 2.41 Å]~\cite{Hansen_2008}  while ramping the field up to 9 T at various temperatures between 1.5 K and 300 K. After each field scan, the magnetic field was set to zero before heating to the next temperature. A NAC powder sample was used as reference for determining the wavelengths and zero offsets. The background from the sample environment was minimized using oscillating radial collimator.

All NPD data were analyzed using the Lebail method,~\cite{LeBail_1987} as implemented in the diffraction analysis package $FullProf$  $Suite$. ~\cite{Rodriguez_1993}  The Rietveld method~\cite{Rietveld_1969} could not be used due to the presence of preferred orientations (unavoidable due the high pressure pelletizing process), whose impact in the integrated intensities could not be satisfactorily described with any of the models available in the $FullProf Suite$ software. The Lebail analysis was thus used to determine the field- and temperature dependence of the lattice parameters and the magnetic propagation vector. The field dependence of the integrated intensities of the magnetic reflection (1/2, 1/2, 1/2) and its incommensurate satellites (1/2, 1/2, 1/2$\pm q$) was determined separately by fitting their diffraction profiles with Gaussian functions. The details of the fitting procedure are presented in the Appendix.

\section{Results}

\subsection{Magnetometry}

As a first step to establish the $H-T$ phase diagram we investigated the temperature dependence of the magnetization $M(T)$ at fixed magnetic fields. Representative curves showing the temperature dependence of the magnetization per unit formula measured using zero field cooled (ZFC) and field cooled (FCC/FCW) protocols are shown in Figs. 1b-d. We attribute the pronounced anomaly observed in the three curves to the colinear-to-spiral transition, an assignment further confirmed by neutron powder diffraction measurements (see next sections). The $T_{spiral}$ value, defined here as the zero of the partial derivative ${\partial M}/{\partial T}$, is $\sim$ 245 K for the curve measured at 1 T. This value is very close to that expected from the chosen synthetic route, and in excellent agreement with the reported value on samples prepared in similar conditions.~\cite{Morin_2016}

The application of a magnetic field has a detrimental effect in $T_{spiral}$, which decreases from $\sim$ 245 K at 1 T to $\sim$ 215 K at 7 T at an approximately linear rate of $\sim$ 8 K / T. We also observe a remarkably different behavior above and below $T_{spiral}$. As shown in Figs. 1b-d, the ZFC, FCC and FCW curves are indistinguishable in the collinear phase ($T > T_{spiral}$). This is not anymore the case below the spiral ordering temperature, where the splitting of the three curves signals the presence of pronounced thermal hysteresis and magnetic irreversibility effects.

The thermal hysteresis affects the FCC (red) and FCW (blue) curves, which bifurcate below $T_{spiral}$ and merge again at a temperature that decreases for increasing magnetic fields. The impact of the field in the broadening of the hysteretic region can be better appreciated in Fig. 1e, showing the difference between both curves $\Delta M_{hyst}$ = $M_{FCC} - M_{FCW}$ at various magnetic fields. The presence of thermal hysteresis, characteristic of first-order phase transitions, is uncommon for a transition between antiferromagnetic phases, and suggests that the collinear and spiral phases coexist in a field-dependent temperature region below $T_{spiral}$.

The existence of magnetic irreversibility below $T_{spiral}$ is revealed by the difference between the curves measured by warming after cooling with- and without field (FCW/ZFC). This effect, more pronounced for high magnetic fields, is illustrated in Fig. 1f, showing the difference between both curves $\Delta M_{irr}$ = $M_{FCW} - M_{ZFC}$  at various magnetic fields. The irreversible magnetization $\Delta M_{irr}$, negligibly small above $T_{spiral}$, rises sharply below this temperature and displays a broad maximum that approximately coincides with the middle point of the hysteretic region. At lower temperatures, in the region without hysteresis in the FCC/FCW curves, $\Delta M_{irr}$ is only weakly temperature dependent. In contrast, it displays a pronounced, non-linear field dependence, illustrated by the $\Delta M_{irr}$ values at selected temperatures shown in Fig. 1g. We note also that the observation of irreversibility \textit{only} below $T_{spiral}$ excludes an origin based in the presence of ferromagnetic impurities, favoring instead an intrinsic origin linked to the presence of the spiral phase.


\begin{figure}[!thb]
\includegraphics[keepaspectratio=true,width=\columnwidth]{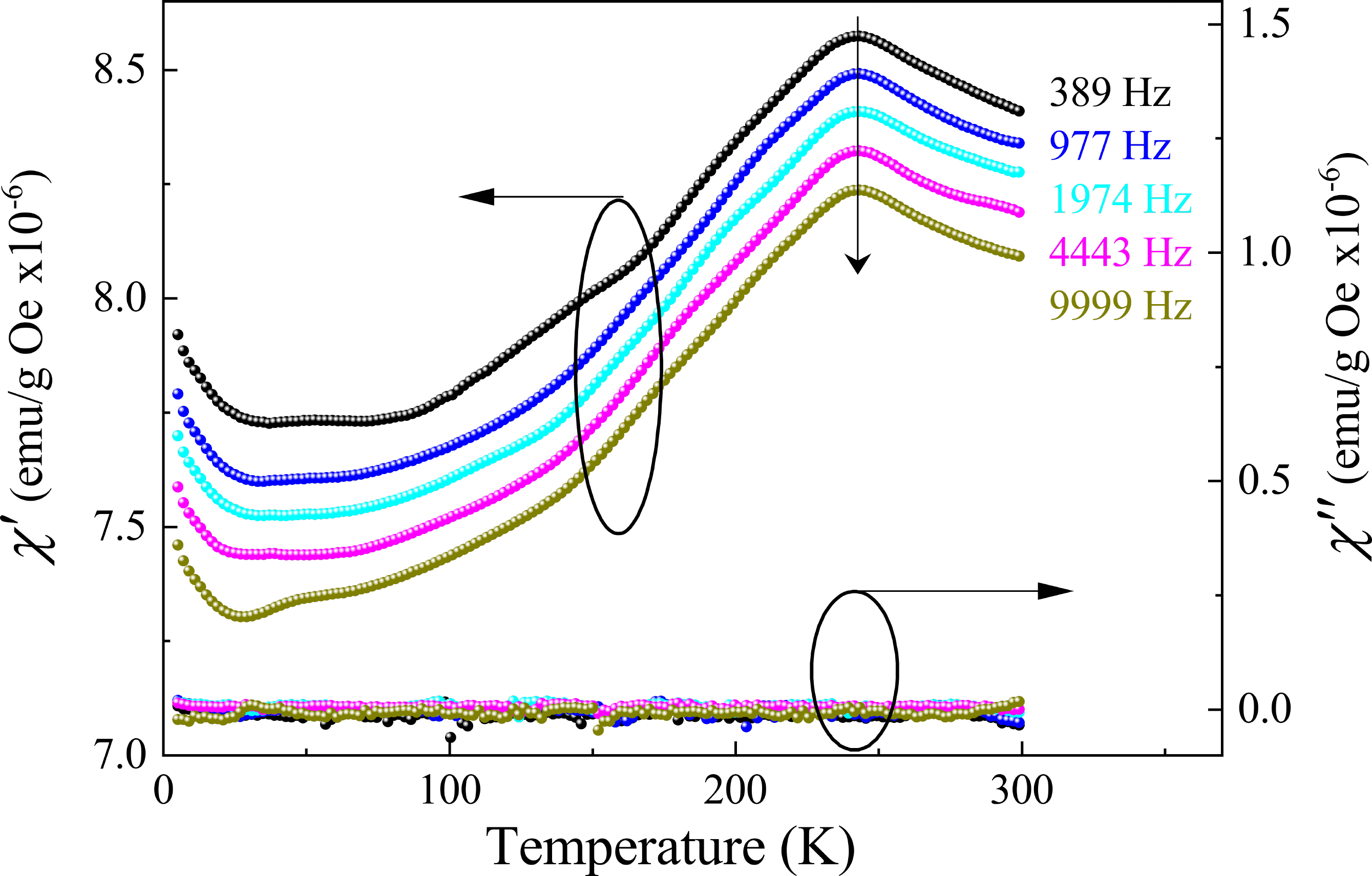}
\vspace{-3mm}
\caption{\label{fig:ac_susc}Temperature dependence of the real and imaginary parts of the $ac$ magnetic susceptibility $\chi'_{ac}(T)$ and $\chi''_{ac}(T)$ measured at different frequencies. The $\chi'_{ac}(T)$ curves are shown equally spaced for clarity.}
\end{figure}

\subsubsection{Weak ferromagnetism} 
Intrinsic irreversibility in frustated antiferromagnets with chemical disorder may have different origins, the most common being either the presence of weak ferromagnetism (WFM), or the existence of a spin-glass state. The main difference between them is the metastability of the spin-glass state, whose response to external stimuli is characterized by complex, time-dependent relaxation phenomena.~\cite{Binder_1986} This has characteristic signatures in frequency-dependent measurements such as the $ac$ susceptibility, where the real part $\chi'_{ac}(T)$ shows a cusp at a temperature $T_f$ with pronounced dependence on the frequency of the $ac$ driving field ($T_f$ is the temperature at which the spin glass frozens at a time scale $\tau$ equal to the inverse of the experimentally used frequency $\omega$ =2$\pi$/$\tau$). The imaginary part $\chi''_{ac}(T)$ displays also a distinctive behavior: it begins to increase above the maximum in $\chi'_{ac}(T)$, has an inflection point which coincides with the peak in $\chi'_{ac}(T)$, and peaks shortly afterwards before dropping down to zero at low temperatures. These phenomena are not observed for magnetic systems in thermodynamic equilibrium, where $\chi''_{ac}(T)$ = 0, and the $\chi'_{ac}(T)$ cusps, if any, are frequency independent.

 To disentangle the two possible origins of the observed irreversibility we measured $\chi_{ac}(T)$ at 6 different frequencies covering approximately 3 decades. As shown in Fig. 2, the only cusp observed in $\chi'_{ac}(T)$ coincides with $T_{spiral}$ and displays no appreciable frequency dependence. Moreover, $\chi''_{ac}(T)$ is featureless and negligibly small in the full temperature range investigated. These observations are at variance with the behavior reported for spin-glass systems,~\cite{Binder_1986,Piatek_2013} and favor a weak ferromagnetic origin for the irreversibility observed below $T_{spiral}$.

\begin{figure*}[!thb]
\includegraphics[keepaspectratio=true,width=180 mm]{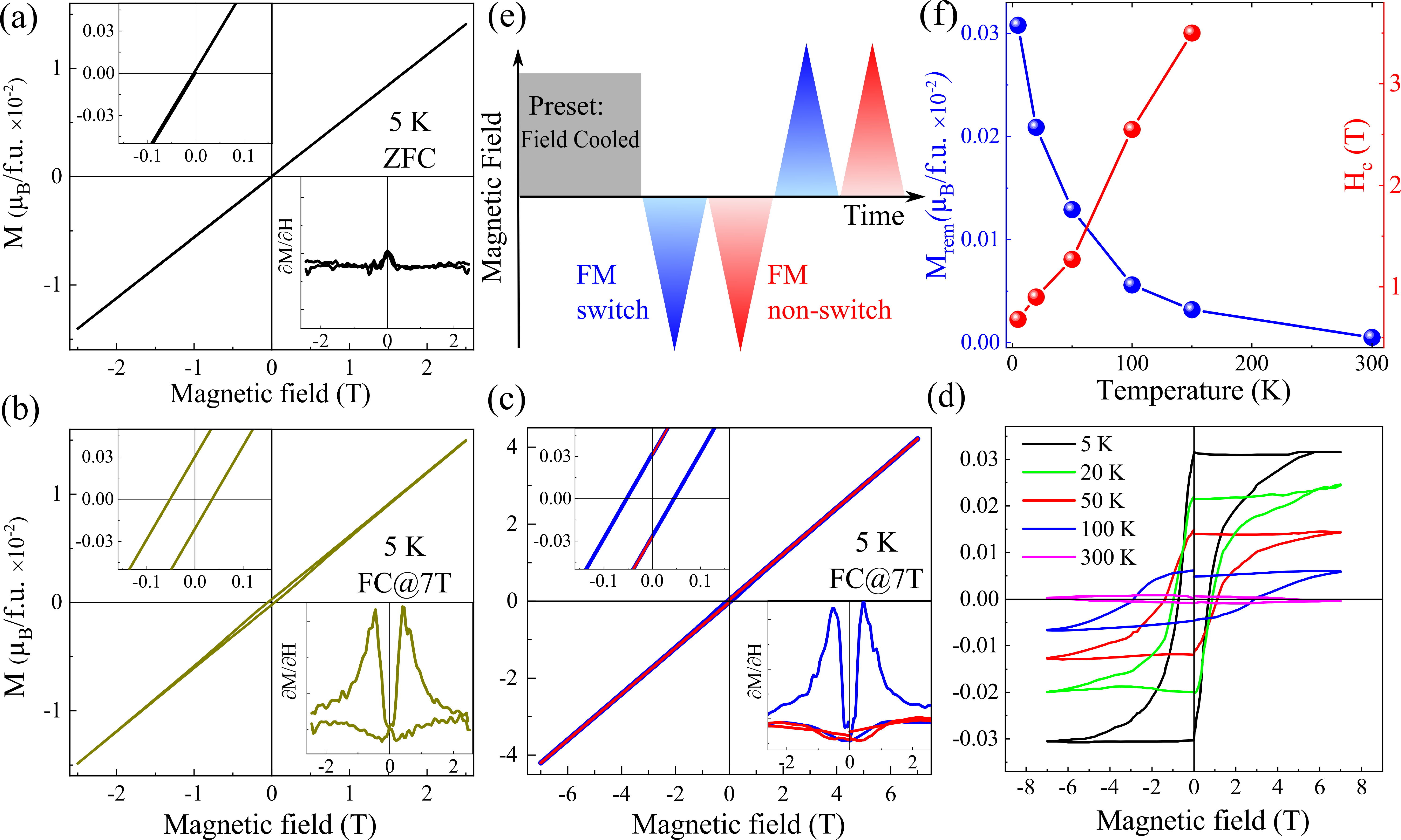}
\vspace{-3mm}
\caption{(a-b) Magnetic field dependence of the magnetization at 5 K during a $M-H$ loop measured after (a) ZFC and (b) FC under magnetic field of 7 T. The up-left inserts show an enlarged view of the low-field region; The down-right inserts show the first derivative ${\partial M}/{\partial H}$ within the same region. (c) Magnetic field dependence of the magnetization at 5 K after FC under 7 T measured using the PUND-like protocol described in the text. The inserts show enlarged views of the low-field region and its first derivative ${\partial M}/{\partial H}$ that, as expected, coincide with those in (b).(d) Switchable part of the magnetization at selected temperatures after subtraction of the non-switchable contribution, as obtained using the PUND protocol schematically shown in (e). Note that the red/blue color code of the different pulses is the same to that in (c). (f) Temperature dependence of the remnant magnetization and the coercive field.}
\label{Fig3:PUND}
\end{figure*}

Further experimental evidence supporting the presence of WFM is provided by the field dependence of the magnetization curves. Fig. 3 shows the $M - H$ cycles measured at 5 K after cooling the sample in zero field (3a) and under 7 T (3b). The ZFC curve presents the linear shape typical of antiferromagnets, indicating that, at this temperature, the WFM domains formed below $T_{spiral}$ are exactly compensated and the applied fields are not large enough to align them. A completely different behavior is observed when the sample is cooled in field. As shown in Fig. 3b, a small hysteresis loop superimposed to the linear antiferromagnetic response is clearly observed. This is better appreciated in Figs. 3c-d, measured using a special protocol similar to the so-called PUND (Positive-Up Negative-Down) method employed to measure electric polarization-electric field hysteresis cycles in ferroelectrics.~\cite{Rabe_2012} This procedure, which allows to separate the switchable and non-switchable components,~\cite{Fina_2011} involves four steps (or "pulses") schematically shown in Fig. 3e. The sample is first cooled under the maximal positive magnetic field (+7 T) down to the desired temperature in order to align the ferromagnetic domains and create a positive magnetization, after which $H$ is set to zero. The field is then ramped to -7 T and back to zero two times. Since the switchable part of $M$ will be reversed during the first "pulse", but not during the second, it can be obtained by subtracting the $M(H)$ curves measured during the first and second field "pulses". The positive-field part can be obtained in a similar way after ramping the field to +7 T and back to zero two times (third and fourth pulses in Fig. 3e).

The switchable part of the magnetization at selected temperatures obtained using this procedure is shown in Fig. 3d. Well defined $M - H$ cycles are observed at $T < T_{spiral}$, with a remnant magnetization that decreases with temperature and virtually disappears in the collinear phase (Fig. 3f). At 5 K the coercive field is $\sim$ 0.7 T, similar to that obtained using the standard procedure (Fig. 3b), and the remnant magnetization is about 3*10$^{-4}$ $\mu_B$/formula unit. This value far below the 6 $\mu_B$ expected from a full alignement of the the Fe$^{3+}$ and Cu$^{2+}$ magnetic moments (5 $\mu_B$ and 1 $\mu_B$ assuming free-ion values), but comparable to the values reported for other weak ferromagnets ~\cite{Mattsson_1994,Li_2012}. This indicates that the uncompensated magnetization arises most likely from a tiny misalignement of the (majoritarily antiferromagnetically coupled) Cu$^{2+}$ and Fe$^{3+}$ moments. An upper limit to the magnetic field $H_{sat}$ needed to reach the fully saturated state can be estimated by extrapolating the linear dependence of the magnetization observed at 20 K (Fig. 4). Assuming the absence of field-induced phase transitions and a saturation value of 6 $\mu_B$ per unit formula we obtain a field close to 1000 T. Although this value is larger than the actual $H_{sat}$, it indicates that the energy scale of the leading magnetic interactions in this material is extremely large. This is consistent with the high ordering temperatures reported for YBaCuFeO$_5$ (425 to 455 K~\cite{Morin_2016,Shang_2018}), the large, negative Curie-Weiss constants indicative of strong, predominantly antiferromagnetic interactions (-790 to -893 K~\cite{Caignaert_1995,Ruiz_1998,Lai_2015}), and with the estimations of the exchange constants from DFT calculations,~\cite{Morin_2015} some of them reaching values as high as $\sim$ 135 meV.

\begin{figure}[!thb]
\includegraphics[keepaspectratio=true,width=\columnwidth]{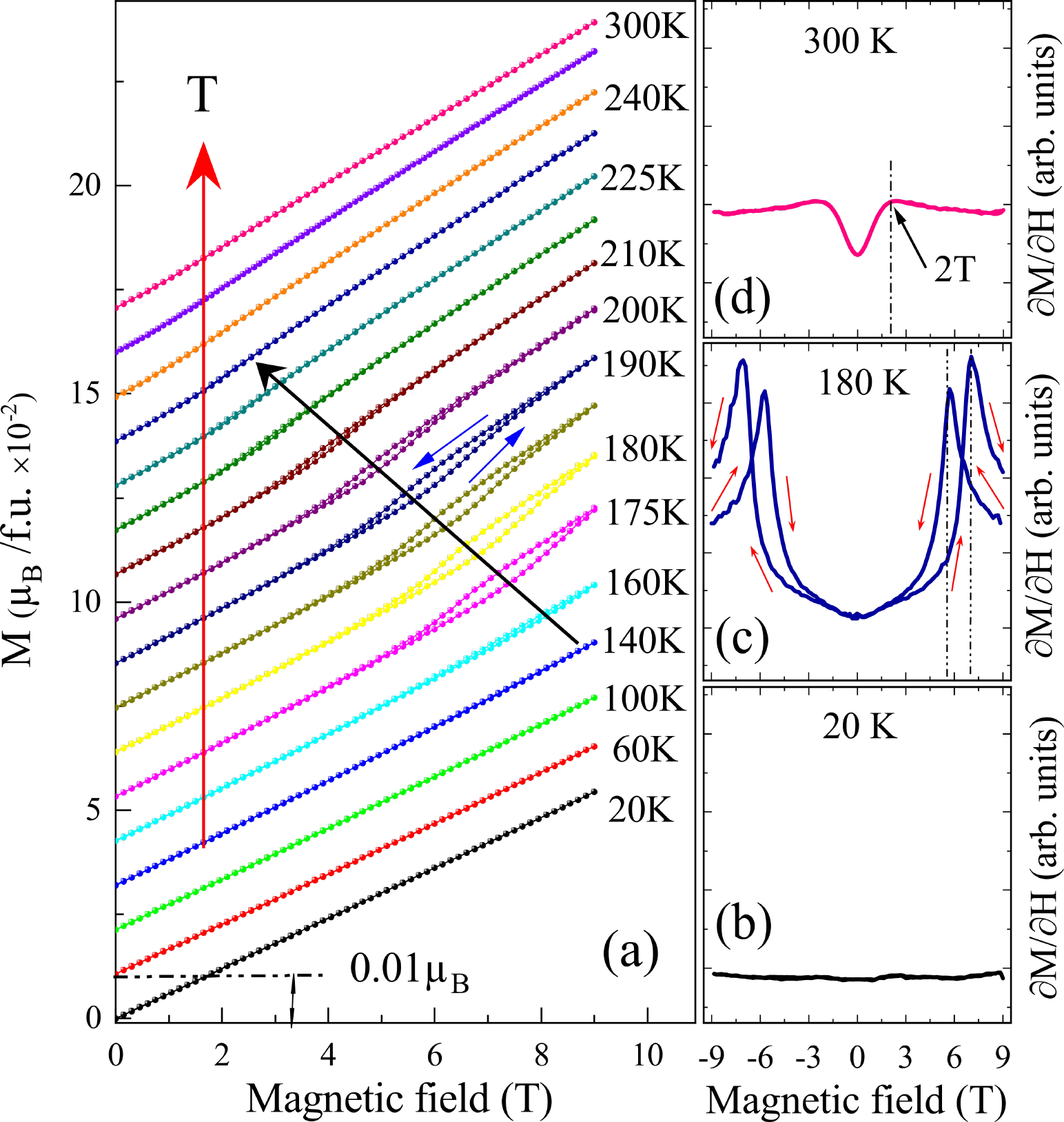}
\vspace{-3mm}
\caption{(a) Magnetic-field dependence of the YBaCuFeO$_5$ magnetization between 0 and 9 T for temperatures between 20 and 300 K. The $M(H)$ curves are offset vertically by 0.01 $\mu_B$/f.u. for clarity. (b-d) Partial derivative curves ${\partial M}/{\partial H}$ at 20, 180 and 300 K through a full hysteresis cycle (-9 T $< H <$ +9 T).}
\label{Fig3:MvsH_pellet}
\end{figure}

 \subsubsection{Field-induced transitions} 
 
Magnetic orders arising from frustrated interactions -such as magnetic spirals - are expected to display a strong response to magnetic fields that may led to the suppression of the spiral state and the stabilization of new magnetic phases ~\cite{Wang_2016,Kindervater_2020}. Because the response of these phases to magnetic fields is usually different from that of the original magnetic order, their emergence can be easily detected by monitoring the slope changes in the isothermal $M(H)$ magnetization curves. For YBaCuFeO$_5$, examples of such curves measured at fields up to 9 T at selected temperatures between 20 and 300 K are shown in Fig. 4a. (see Fig. 11 in the Appendix for an extended data set). Before starting the measurements the sample was first zero-field cooled down to 20 K. The magnetic field was then increased in steps up to 9 T, and set to zero before heating to the next temperature. 

At 20 K, we note that $M$ is a linear function of $H$ without any clear inflection point. This is better appreciated in the ${\partial M}/{\partial H}$ derivative, which is constant within the investigated $H$ range (Fig. 4b). The $M(H)$ linearity subsists up to $\sim$ 140 K, but between this temperature and $T_{spiral}$ a slope change with hysteresis indicative of a first-order transition is clearly observed. The transition is particularly evident in the first derivative curves, exemplified in Fig. 4c by the data measured at 180 K through a full $M(H)$ hysteresis cycle. We note that both, the magnetic field $H_1$ corresponding maximum of ${\partial M}/{\partial H}$ and the width of the hysteretic region decrease by approaching $T_{spiral}$ $\sim$ 245 K. Above this temperature and up to 300 K we still observe a slope change, albeit less pronounced, with a maximum at $H_2$ = 2 T (Fig. 3d see also Fig. 12 in the Appendix for an extended data set). Contrarily to $H_1$, $H_2$ is temperature independent and shows no hysteresis, suggesting that slope change at 2 T corresponds to a distinct, possibly second-order field-induced transition.

\subsection{Powder Neutron Diffraction}

To get insight on the nature of the field-induced magnetic phases we conducted NPD measurements under magnetic field. One should nevertheless mention that the interpretation is in general not as straightforward as in the case of a single crystal. This is because the field-induced magnetic configuration of a particular grain will strongly depend on the grain orientation with respect to the external magnetic field, which is in turn correlated with the geometry of the diffraction process. Moreover, the use of a sintered rod, necessary to avoid the movement of the grains under the action of the magnetic field, results in the existence of preferential orientations that makes a quantitative analysis of the magnetic Bragg reflection intensities extremely difficult. For these reasons we discuss here exclusively the evolution of two variables. On one side, the position of the magnetic Bragg reflections, which provides information about the field-dependence of the magnetic propagation vector. On the other, the field-dependence of the magnetic intensities relative to that of the zero-field patterns, where the observation of a clear discontinuity at $\sim$ 2 T provides additional support to the existence of the field-induced transition inferred from the magnetization data.

\begin{figure}[!bht]
\includegraphics[keepaspectratio=true,width=\columnwidth]{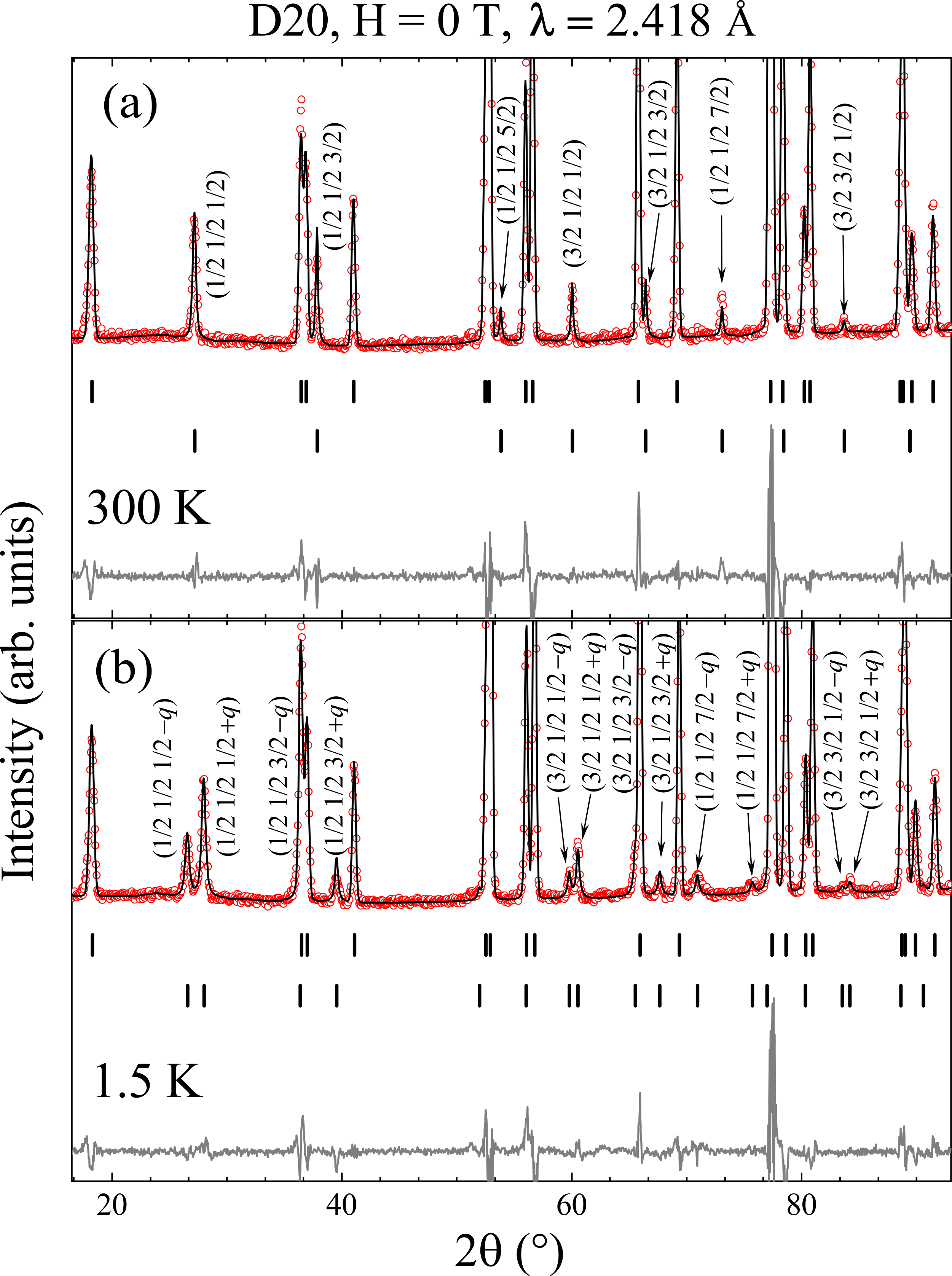}
\vspace{-3mm}
\caption{NPD pattern measured at 0 T on D20 at 300 K (a) and 1.5 K (b). Red circles: experimental data. Black lines: Lebail fits. The upper row of vertical ticks indicates the positions of the nuclear Bragg reflections allowed by the space group $P4mm$. In (a), the lower row indicates the positions of magnetic reflections associated to the collinear magnetic structure with \emph{\textbf{k$_c$}} = (1/2, 1/2, 1/2); in (b), those associated to the spiral magnetic structure with \emph{\textbf{k$_i$}} = (1/2, 1/2, 1/2$\pm q$). The grey lines are the difference curves between observed and calculated diffraction patterns.}
\label{Fig4:NPD_refine}
\end{figure}

\subsubsection{PND at zero field}

Figs. 5a and 5b show the zero-field NPD patterns at 300 K (collinear phase) and 1.5 K (spiral phase). In both patterns, the nuclear Bragg reflections can be indexed with a tetragonal cell of $P4mm$ symmetry. At RT, the reflections of magnetic origin correspond to the propagation vector \textbf{k$_c$} = (1/2, 1/2, 1/2), and at 1.5 K, to an incommensurate modulation with \textbf{k$_i$} = (1/2, 1/2, 1/2$\pm q$) as reported in previous works ~\cite{Morin_2015,Morin_2016, Shang_2018}. At 1.5 K the $q$ value is $\sim$ 0.115, very close of the value expected from the universal law relating $q$ and $T_{spiral}$ ($q$ $\sim$ 0.117)~\cite{Shang_2018,Scaramucci_2020}. The temperature dependence of the (1/2, 1/2, 1/2) and (1/2, 1/2, 1/2$\pm q$) magnetic reflections, their integrated intensities, and the magnetic modulation vector $q$ are shown in Figs. 6a-c. In the spiral phase, $q$ and the integrated intensities of the magnetic satellites decrease with temperature and vanish above $T_{spiral}$, where they are replaced by the commensurate reflection (1/2, 1/2, 1/2). As expected, the $T_{spiral}$ value ($\sim$ 250 K) measured at $H$ = 0 and defined here as the maximum of the (1/2, 1/2, 1/2) reflection is slightly higher than the value inferred from the $M(T)$ measurements at 1 T ($\sim$ 245 K, Fig. 1b ). As in previous works,~\cite{Morin_2015,Morin_2016, Shang_2018} we observe here phase coexistence at low temperatures, which is quite significant between $T_{spiral}$ and $\sim$ 180 K. Below this temperature the fraction of collinear phase becomes nearly constant and very close to the detection limit. However, it is worth mentioning that the wavelength used in this study makes difficult to extract small values of the (1/2, 1/2, 1/2) integrated intensity due to the superposition with the tails of the incommensurate satellites (see Appendix for details on the fitting procedure). The nonzero values obtained below 200 K should be thus considered with caution. Interestingly, the point where the temperature dependence of the (1/2, 1/2, 1/2) integrated intensity starts to deviate from linearity ($\sim$ 180 K) is reasonably close to the lower phase coexistence limit inferred from magnetometry. Hence, we use here this simple criterion to define the lower phase coexistence boundary from NPD data. Measurements with a longer wavelength will be however be necessary to establish unambiguously the actual phase coexistence limit in our sample.   

\begin{figure}[!tbh]
	\includegraphics[keepaspectratio=true,width=\columnwidth]{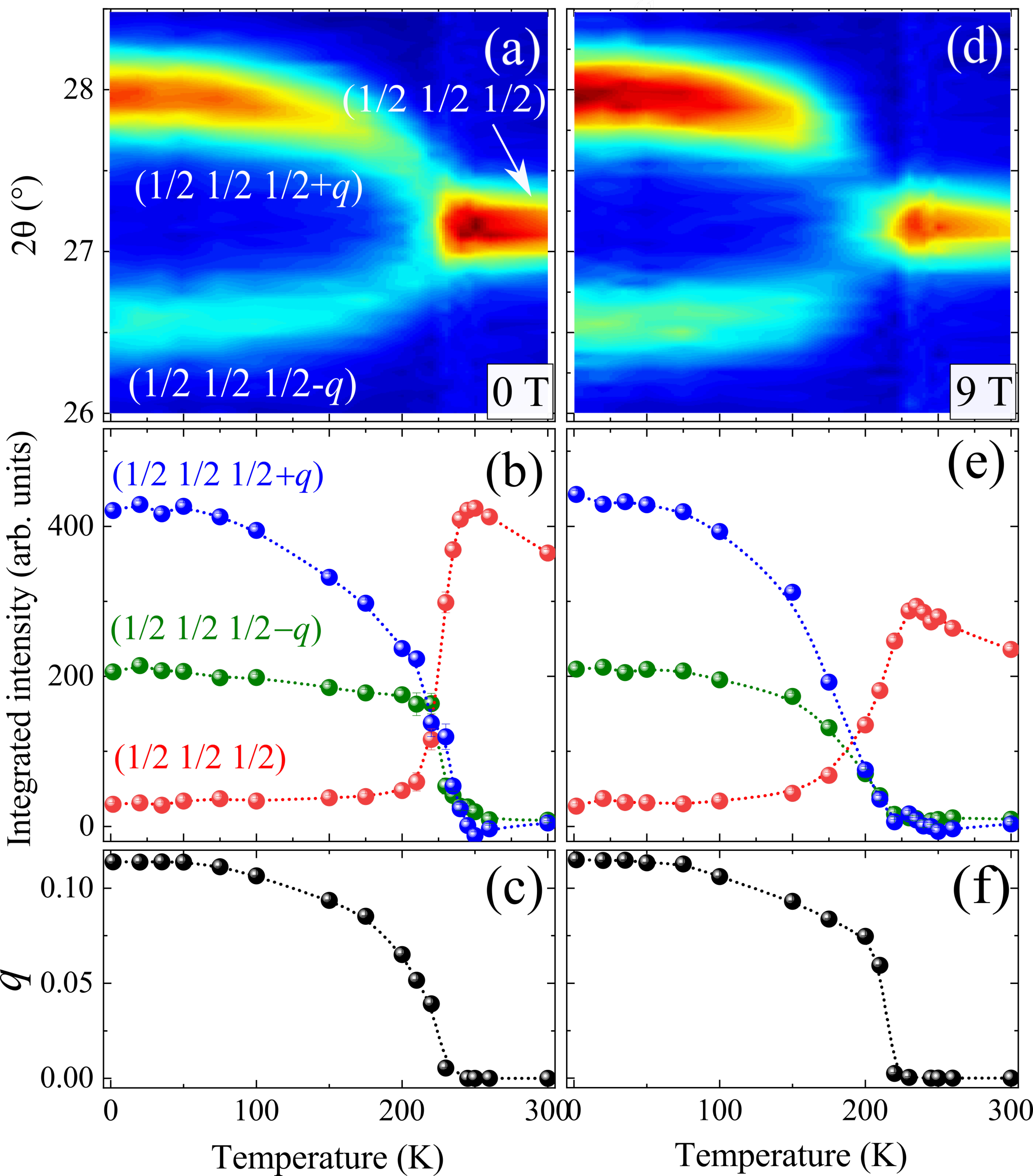}
	\vspace{-3mm}
	\caption{\label{fig:2D_plot} 2D contour plots showing the temperature dependence of the (1/2, 1/2, 1/2) and  (1/2, 1/2, 1/2$\pm q$) magnetic reflections measures under 0 (a) and 9 T (d). The temperature dependence of their integrated intensities and the magnetic modulation vector $q$ are shown in (b-c) at 0 T, and (e-f) at 9 T.}
	\vspace{-1mm}
\end{figure}

\begin{figure}[!bht]
\includegraphics[keepaspectratio=true,width=\columnwidth]{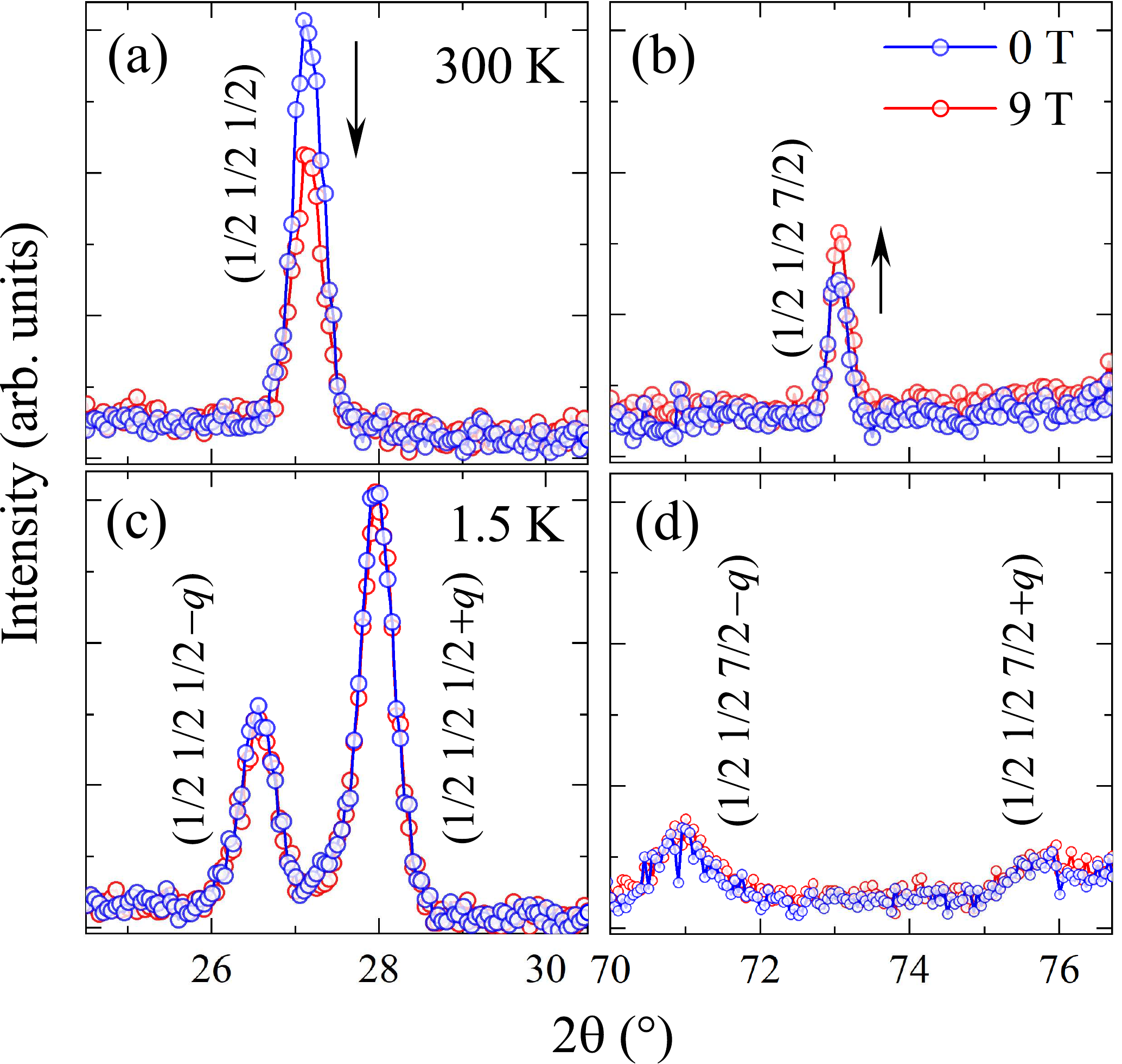}
\vspace{-3mm}
\caption{Comparison between the intensities of selected magnetic reflections measured at 0 (blue dots) and 9 T (red dots). (a-b) In the collinear phase at 300 K. (c-d) In the spiral phase at 1.5 K.}
\label{Fig6:NPD_comparison}
\end{figure}

\begin{figure*}[!bht]
\includegraphics[keepaspectratio=true,width=180 mm]{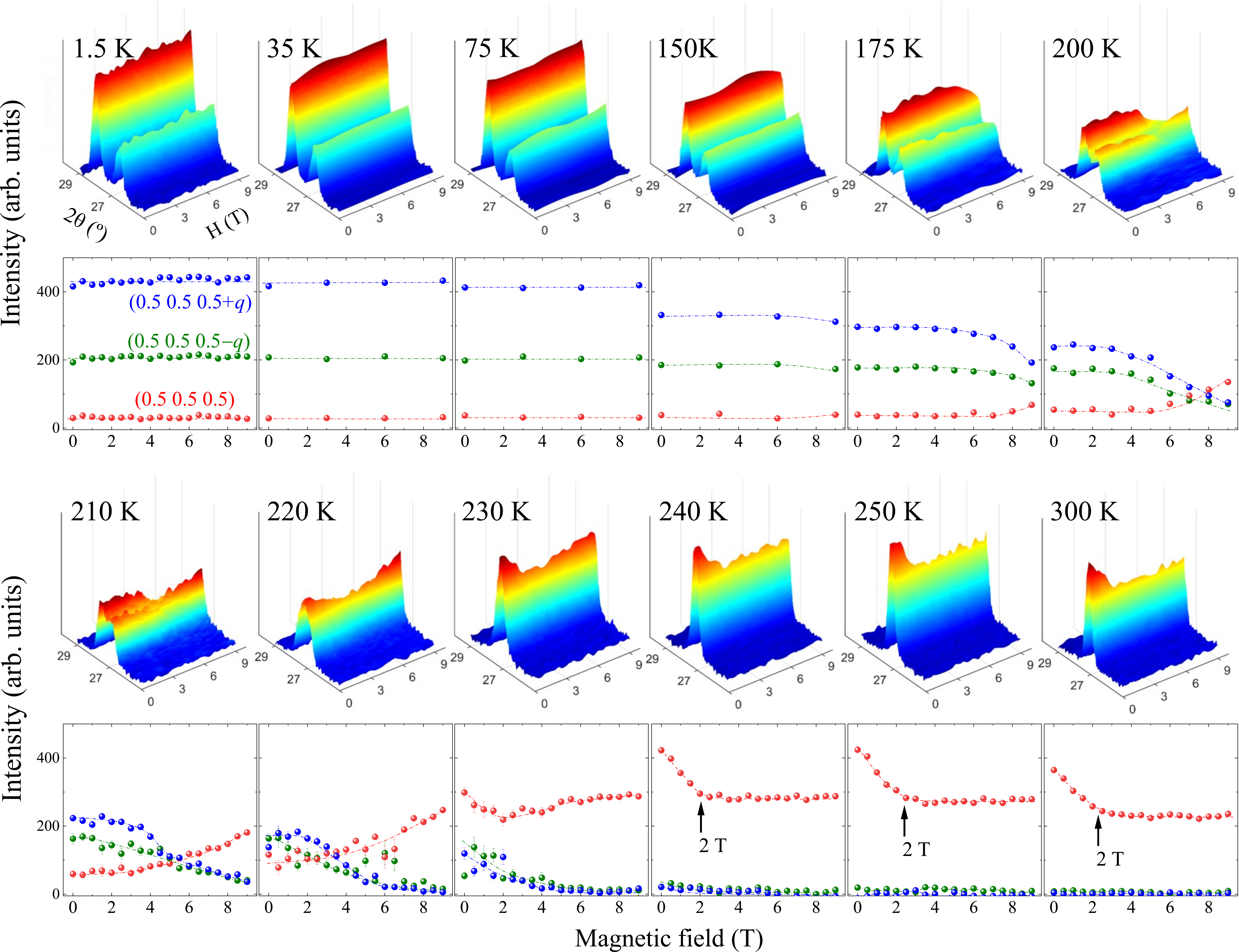}
\vspace{-3mm}
\caption{Contour maps showing the field dependence of the magnetic Bragg reflection (1/2, 1/2, 1/2) associated to the high-temperature collinear phase and the (1/2, 1/2, 1/2$\pm q$) satellites of the low-temperature magnetic spiral phase at selected temperatures together with the field dependence of their integrated intensities.}
\label{Fig5:NPD_evolution}
\end{figure*}

\subsubsection{PND at fields up to 9 T}

Figs. 6d-f show the temperature dependence of the same 2$\theta$ region shown in Figs. 6a-c after applying a magnetic field of 9 T. At low temperatures ($T$ $<$ $T_{spiral}$) we still observe an incommensurate modulation with $q$ and magnetic satellite intensities nearly identical to those observed at 0 T. The only noticeable modifications with respect to the zero field data are a slightly different temperature dependencies of both, $q$ and the satellite's integrated intensities, a decrease of $T_{spiral}$, and a much larger temperature region where the spiral and collinear phases coexist in sizable amounts. Using the criterion described in the previous section, the lower boundary of the phase coexistence region at 9 T will be $\sim$ 100 K, which is again consistent with the value inferred from magnetometry. We do not observe any signature of the field-induced WFM component, which would appear as an extra contribution to the nuclear reflections in the NPD data. This is an expected finding given that the observation of WFM requires a FC process, and the NPD data were collected after cooling the sample in zero field. For data obtained after a FC process, the small value of the net magnetization (3*10$^{-4}$ $\mu_B$/formula unit at 7 T), far below the detection limit of the technique ($\sim$ 0.1 $\mu_B$), will anyway prevent the observation of the WFM signal using the NPD technique.

At higher temperatures ($T$ $>$ $T_{spiral}$) we note that the integrated intensity of the (1/2, 1/2, 1/2) reflection takes values significantly lower than those observed in absence on magnetic field. This is better appreciated in Fig. 7, showing selected regions of the NPD patterns measured at 1.5 and 300 K under 0 and 9 T. In contrast to the magnetic field invariance of the reflections associated to the spiral phase at 1.5 K, the figure reveals clear intensity differences in the magnetic reflections associated to the commensurate propagation vector \textbf{k$_c$} at RT. More precisely, the intensity of the ($h$+1/2, $k$+1/2, $l$+1/2) reflections with $l$ = 0 decreases, and that of those with $l$ $\neq$ 0 increases. This suggests the existence of a different magnetic phase at 9 T described by the same propagation vector \textbf{k$_c$} = (1/2, 1/2, 1/2) observed in absence of magnetic field.

The different impact of the magnetic field in the spiral and collinear phases is better appreciated in Fig. 8, showing the field dependence of the (1/2, 1/2, 1/2$\pm q$) and (1/2, 1/2, 1/2) reflections and their integrated intensities at several temperatures between 1.5 and 300 K. We note that the invariance of the incommensurate satellites for fields up to 9 T observed at 1.5 K is preserved up to $\sim$ 140 K.  Above this temperature, and beyond a critical field, modifications of the magnetic reflections associated to the spiral phase become evident. The 2$\theta$ positions of the incommensurate satellites become closer, their intensities decrease, and the commensurate reflection (1/2, 1/2, 1/2) suddenly appears. The impact of magnetic field on the spiral phase is thus similar to that of temperature, with the spiral fully disappearing beyond a critical field $H_{1}(T)$ that decreases by increasing temperature and vanishes at $\sim$ 250 K. Phase coexistence is observed in a temperature region that becomes narrower by approaching the zero-field  $T_{spiral}$ value, in agreement with the magnetization results.

Between 250 and 300 K the propagation vector is commensurate for any field between 0 and 9 T. However, in contrast with the behavior observed in the spiral phase, the intensity of the (1/2, 1/2, 1/2) reflection decreases continuously with $H$ up to $\sim$ 2 T. Above this value it remains constant up to the largest field investigated. This behavior confirms the existence of a field-induced, second-order transition at $H_2$ $\sim$ 2 T, in agreement with the conclusion inferred from the field dependence of the isothermal magnetization curves.

As mentioned before, a quantitative analysis of the integrated intensities cannot be satisfactorily performed with the present data. However, the field-dependence of the Bragg reflections above 250 K can already give some hints on the modifications of the zero-field commensurate magnetic order under the action of a magnetic field. On one side, the preservation of the commensurate propagation vector \textbf{k$_c$} above and below $H_2$ $\sim$ 2 T demonstrates that the periodicity of the low (COL1) and high-field (COL2) magnetic arrangements is identical. On the other, the COL1 and COL2 phases feature the same reflection set, albeit with different intensities. Moreover, the small value of the critical field (2 T) is indicative of a low magnetic energy scale. According to DFT calculations, the smallest nearest-neighbor (NN) couplings in YBaCuFeO$_5$ are those along the \textbf{c} axis \cite{Morin_2015}. Moreover, the magnetic anisotropy energy is probably comparable in order of magnitude \cite{Schron_2012}. Possible changes in the magnetic arrangement compatible with these observations could thus include a spin-flop transition, or a change in the coupling schema along the \textbf{c} axis (as shown in  \cite{Morin_2015}, at least three different magnetic configurations are consistent with \textbf{k$_c$} = (1/2, 1/2, 1/2) and the observed reflection set). Due to the use of a ceramic rod it is extremely difficult to interpret the response of the magnetic intensities to an external magnetic field shown in Fig. 7 as signature of one of these possibilities. However, the large in-plane NN couplings in YBaCuFeO$_5$ and the low anisotropy of the collinear phase~\cite{Lai_2017} led us to speculate about the possibility of a spin-flop transition. Such transitions involve a change in the orientation of the antiferromagnetic configuration from parallel to transverse when the external magnetic field is applied along an easy antiferromagnetic direction, and are observed in antiferromagnets with anisotropy energies significantly weaker than the exchange interactions.

In YBaCuFeO$_5$ single crystals with $T_{spiral}$ = 160 K, the magnetization was reported to have an easy plane perpendicular to the fourfold axis parallel to \textbf{c} at any temperature below $T_{collinear}$ $\sim$ 440 K.~\cite{Lai_2017} This is in agreement with refs.~\cite{Morin_2016,Shang_2018}, where lower $T_{spiral}$ values were found to result in inclined spirals with the rotation plane closer to the crystallographic \textbf{ab} plane. For small values of the magnetic anisotropy, a magnetic field applied along \textbf{a} or \textbf{b} could thus induce a spin-flop transition and align the magnetic moments parallel to the \textbf{c} axis. Interestingly, the anisotropy of the collinear phase was found to be quite modest whereas that of the spiral phase was significantly larger \cite{Lai_2017}. This could explain the stability of the spiral under application of a magnetic field, and supports the observation of a spin-flop transition $only$ in the collinear phase at a modest magnetic field $H_2$ = 2 T. This field is indeed much smaller than the extrapolated $H_{sat}$  $\sim$ 1000 T, which reflects the overall energy scale of the exchange couplings in YBaCuFeO$_5$. This, together with the smooth evolution of the magnetic reflections for fields up to $H_2$, makes the spin-flop scenario reasonably plausible. Magnetization and neutron diffraction measurements on single crystals will be nevertheless necessary to establish whether it is realized in YBaCuFeO$_5$.

\begin{figure}[!bht]
	\includegraphics[keepaspectratio=true,width=\columnwidth]{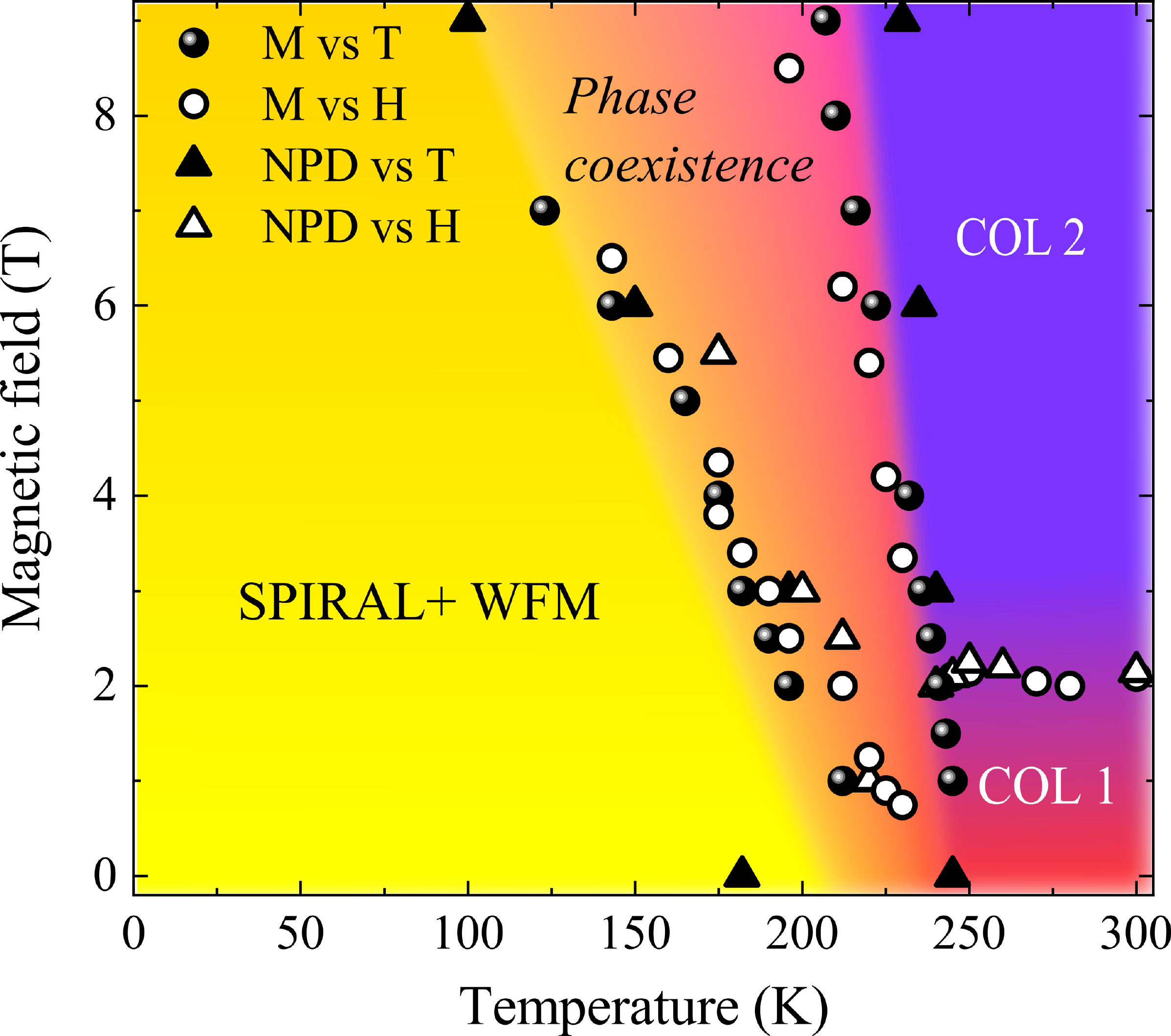}
	\vspace{-3mm}
	\caption{\label{fig:phase_diagram}Magnetic phase diagram of polycrystalline YBaCuFeO$_5$. The techniques used to determine the lines separting the different magnetic phases are indicated in the legend.}
	\vspace{-3mm}
\end{figure}

\subsection{Phase diagram}

Based on the data presented in previous sections, we can now establish the magnetic phase diagram for polycrystalline YBaCuFeO$_5$ (Fig. 9). Phase lines constructed from the maximum of the (1/2, 1/2, 1/2) reflection in the NPD versus temperature data and the point where its integrated intensity starts to deviate from linearity (black triangles), the setup/anomaly at 2 T of this reflection in the NPD versus field data (white triangles), the limits of the hysteresis region in the FCC - FCW curves (black circles), and the $M(H)$ curves (white circles) mark the boundaries of four different regions: an incommensurate spiral phase with propagation vector \textbf{k$_i$} and a small WFM component, two different commensurate phases COL1 and COL2 with identical propagation vector \textbf{k$_c$}, and a commensurate-incommensurate phase coexistence region whose thermal stability  increases with the applied field. This region, where hysteresis is observed as a function of both temperature and magnetic field, extends along the line separating the incommensurate spiral phase from the two commensurate phases COL1 and COL2. The two spiral $\rightarrow$ COL1 and spiral $\rightarrow$ COL2 transitions in YBaCuFeO$_5$ present thus some characteristics typical from 1$^{st}$ order transitions in spite of the smooth temperature and magnetic field dependence of the magnetic modulation vector $q$ (Figs. 6 and 8). This unusual behavior, not reported for other high-temperature spiral magnets such as CuO ~\cite{Wang_2016}, could be a consequence of the disorder-based mechanism at the origin of the spiral. Whereas in CuO the magnetic frustration arises from the classical competition between NN and NNN exchange interactions, the spiral state in YBaCuFeO$_5$ originates around a few, randomly distributed Fe-Fe frustrating bonds, which induce a canting of the Cu/Fe magnetic moments around them refs.~\cite{Scaramucci_2018,Scaramucci_2020}. In the collinear phase, such "impurity" bonds could play a role similar to nucleation centers in 1$^{st}$ order transitions, which grow at expenses of the preexisting phase leading to the phase coexistence and hysteresis characteristic of this kind of transition. The COL1 $\rightarrow$ COL2 transition at 2 T displays in contrast no hysteresis, as expected in a prototypical 2$^{nd}$ order phase transition.

As shown in Fig. 9, the phase coexistence region broadens for higher fields and lower temperatures. The field dependence of the spiral nucleation and growth rate at different temperatures have not been investigated theoretically. However, we can speculate that an external magnetic field could modify the canting angle distribution of the Fe/Cu magnetic moments around the Fe-Fe impurities necessary for the spiral stability ~\cite{Scaramucci_2018,Scaramucci_2020}. The observation of an enlarged phase coexistence region at higher magnetic fields and lower temperatures also suggest that a decrease of the thermal fluctuations may be necessary for the spiral to become majoritary and ultimately replace the collinear phase. Finding whether this is the case will however require additional theoretical work. At a purely experimental level, linear extrapolation of the phase coexistence boundaries of the phase diagram suggests that, at base temperature, the (pure) spiral + WFM and COL2 phases could exist, respectively, for magnetic fields lower than $\sim$ 18-20 T, and larger than 58-60 T. Magnetization and neutron diffraction experiments will be however necessary to check this prediction. These measurements will be also needed  to explore the higher-field part of the phase diagram, which, given the large value of the estimated $H_{sat}$ upper limit  ($\sim$ 1000 T) at low temperatures, gives plenty of room to observe additional field-induced transitions.

\subsection{Perspectives for spiral orientation and polarization manipulation}

An unexpected finding uncovered by the magnetic phase diagram is the observation of a field-driven transition $only$ in the collinear phase. In the yellow region Fig. 9, our NPD data did not reveal any signature suggesting a modification of the envolvent shape and/or orientatiom of the spiral, \textit{a priori} more prone to be destabilized by a magnetic field. Interestingly, a similar behavior has been observed in CuO, another easy-plane, high temperature spiral multiferroic. As shown by Wang et al. ~\cite{Wang_2016}, the magnetic field needed to suppress the multiferroic spiral phase observed between 213 and 230 K is temperature-dependent and can reach values as high as 50 T. In contrast, the collinear phase oberved below 213 K undergoes a spin-flop transition at a much lower, temperature independent field (10.5 T). These findings have been explained as the result of the fine balance between the exchange interactions and the anisotropy energies associated to the easy-plane and the hard axis. Since the last ones are temperature dependent and differ by a factor of $\sim$36 at low temperatures, a change in the orientation of the magnetic moments with respect to the magnetic field can substantially change the energy balance. A similar scenario could also be pertinent for YBaCuFeO$_5$, but measurements of the anisotropy energy in single crystals and information on the phase diagram along the different crystal axes will be necessary to model the stability ranges of the different magnetic phases.

Another intriguing observation inferred from our data is the emergence of a WFM component \textit{only} in the spiral phase under the application of a magnetic field. It is worth mentioning that ferromagnetism is symmetry-forbidden (and indeed not observed) for $H$ = 0~\cite{Morin_2015}. However, its existence for $H$ $>$ 0 is unambiguously demonstrated by our magnetization and $ac$ susceptibility data. The origin of the WFM component is unclear, but the fact that it is only observed in the spiral phase suggests a common origin for both phenomena. A possible scenario could be that the canting of the Cu/Fe magnetic moments in the vicinity of a Fe-Fe impurity ~\cite{Scaramucci_2018,Scaramucci_2020} is not exactly compensated when an external magnetic field is applied. Probing whether this is the case is out of the scope of this study and  will require additional theoretical work.

A third point needing further investigations concerns the orientation of the WFM component with respect to the spiral rotation plane, and whether both orientations are linked. This is a very important information because it could open the way towards the manipulation the spiral orientation (and hence the polarization direction). As shown in Fig. 3d, the WFM component can be switched with relatively modest fields ($\sim$ 0.7 T at 5 K), but only after a field-cooling process. If the orientations of the WFM component and the spiral are linked, it seems reasonable to expect that a magnetization reversal after field-cooling could result also in a modification of the spiral orientation. Polarization and neutron diffraction measurements under magnetic field on single crystals will be necessary to establish whether this is the case in YBaCuFeO$_5$.

\section{Summary and conclusions}

To conclude, we report here the first $H$-$T$ phase diagram of polycrystalline YBaCuFeO$_5$, a high temperature spiral magnet and multiferroic candidate. Using magnetometry and neutron powder diffraction at low temperatures (2-300 K) and high magnetic fields (0-9 T) we reveal the existence of an incommesurate spiral phase, two collinear phases, and a broad phase coexistence region separating the phases with commensurate and incommensurate magnetic orders. 

Contrarily to our initial expectations, we observe a field-induced transition in the high-temperature collinear phase at $H_2$ = 2 T without change of propagation vector. In contrast, the spiral phase appears to be stable up to much higher fields (in particular at low temperatures) when the sample is cooled in zero field. Another unexpected finding of our study is the observation of weak ferromagnetism coexisting with the spiral modulation. Moreover, the WFM component can be switched by a modest magnetic field if the sample is field-cooled below $T_{spiral}$. Our data could not provide information about the existence of a coupling between the orientation of the spiral and the WFM component. However, given that the last one appears exclusively in the spiral phase, it seems reasonable to assume that some degree of coupling exists. If this is the case, the tunability of the WFM component could open the way to a magnetic field-driven modification of the spiral orientation, directly related with the polarization direction. Moreover, an electric field acting on the polarization could also modify the magnetization. This would open new perspectives for a low-energy control of $M$ and $P$, and the possible use of YBaCuBaO$_5$ in technological applications based in the magnetoelectric effect.

\section{Acknowledgements}

 We thank T. Hansen and E. Leli\`{e}vre for technical support during the neutron powder diffraction measurements under magnetic field, and M. M\"{u}ller for stimulating discussions. The allocation of neutron beamtime at the Institut Laue Langevin, (Grenoble, France), and the financial support from the Swiss National Science Foundation (Grants No. 200021-141334/1 and No. 206021-139082) are gratefully acknowledged. 
 
\section{Appendix}

\subsection{Integrated intensities of the low 2$\theta$ magnetic reflections}

The integrated intensities of the commensurate (1/2, 1/2, 1/2) reflection and the incommensurate (1/2, 1/2, 1/2$\pm q$) satellites shown in Figs. 6 and 8 were determined by fitting the neutron diffraction profiles with three Gaussian functions plus a linear background between 23$^{\circ}$  $\leq$ 2$\theta$ $\leq$ 32$^{\circ}$. The free parameters were the 2$\theta$ positions and maximal intensities for the three gaussians, plus two additional parameters describing a linear background. The fits were performed in sequential mode, starting from 1.5 K upwards.  Because of the progressive merging of the three reflections by approaching $T_{spiral}$, the widths of the three reflections could not be refined independently. Hence, we fixed them to the values obtained at temperatures as far as possible from $T_{spiral}$ (1.5 K for the spiral phase, 300 K for the collinear phase). This assumption is most probably not valid close to the spiral ordering temperature, where broader reflections are expected due to the phase coexistence and hence smaller collinear/spiral diffraction domains. However, it was the only way to avoid the divergence of the fits by approaching $T_{spiral}$. Examples of the fits at selected temperatures for $H$ = 0 and 9 T are presented in Fig. 10, with the contributions of the three reflections to the full profile displayed in different colors.

\begin{figure}[!hbt]
\includegraphics[keepaspectratio=true,width=\columnwidth]{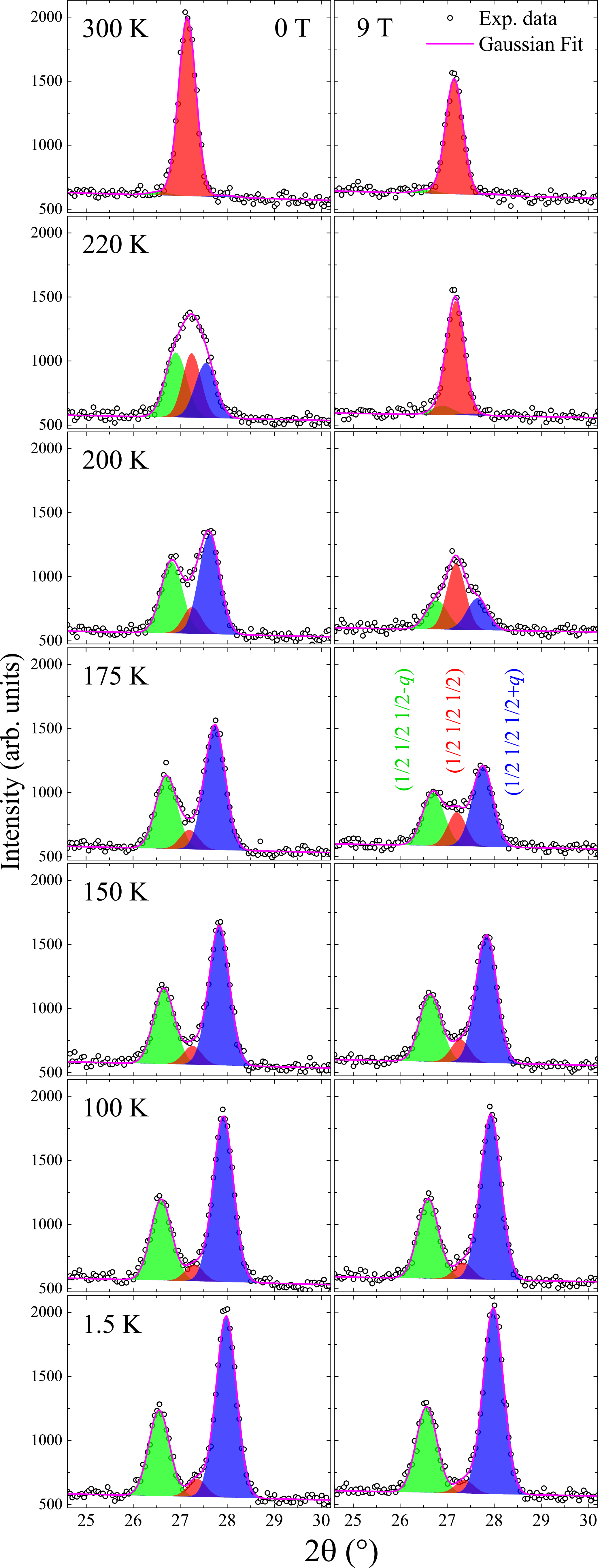}
\vspace{-3mm}
\caption{Temperature dependence of the commensurate (1/2, 1/2, 1/2) reflection and the incommensurate (1/2, 1/2, 1/2$\pm q$) satellites measured at  0 (left column) and 9T (right column). The powder neutron diffraction profiles have been fitted with three gaussian functions as described in the text. Black circles: experimental data. Pink line: fitted profile. Green, red and blue shaded areas: integrated intensities of the (1/2, 1/2, 1/2$-q$), (1/2, 1/2, 1/2) and (1/2, 1/2, 1/2$+q$) magnetic reflections, respectively.}
\label{Fig:Gaussian Fitting of integrated peaks}
\end{figure}

\subsection{Extended magnetization data}

As complement to Fig. 4, an extended set of magnetization curves and their first derivatives is presented in Figs. 11 and 12. Fig. 11 shows the magnetic field dependence of the magnetization in YBaCuFeO$_5$ at 20 different temperatures between 20 and 300 K through a full $M(H)$ hysteresis cycle (-9 $< H > $ +9 T). The first derivative ${\partial M}/{\partial H}$ of these curves is shown Fig. 12. 

For temperatures between 20 and $\sim$ 140 K the magnetization increases in an almost linear way with $H$. In contrast, a pronounced non-linear behavior accompanied by hysteresis is clearly observed between $\sim$ 140 K and $T_{spiral}$. The presence of hysteresis can be particularly well appreciated in the derivative curves of Fig. 12, where the anomaly associated to the spiral to collinear transition appears as a sharp, field and temperature-dependent maximum.

As shown in Fig. 11, the $M(H)$ curves are also non-linear between $T_{spiral}$ and 300 K. This can be better appreciated in Fig. 12, where the first derivative signals the presence of a slope change at a temperature-independent field $H$ = 2 T. The associated anomaly is less sharp than in the case of the spiral-collinear transition and it does not show hysteresis, suggesting the presence of a field-induced transition of different nature. This was further confirmed by the field-dependent PND measurements, which identified the anomaly with a transition between two different collinear phases COL1 and COL2 with identical  magnetic propagation vector.

\begin{figure*}[!hbt]
\includegraphics[keepaspectratio=true,width=180 mm]{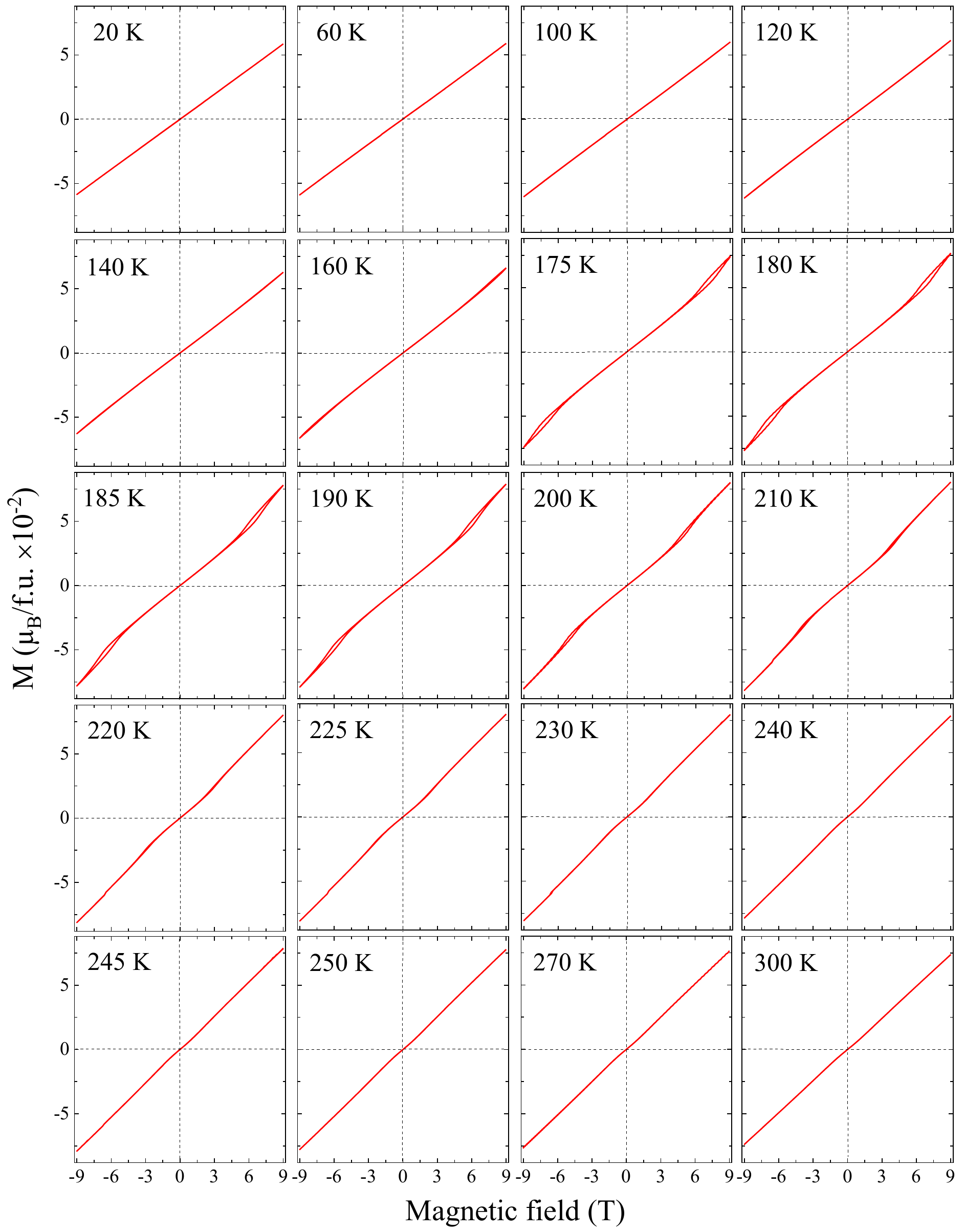}
\vspace{-3mm}
\caption{Magnetic-field dependence of the magnetization $M$ measured along a full -9T to 9T loop for polycrystalline YBaCuFeO$_5$ at selected temperatures between 20 and 300 K.}
\label{Fig:MvsH}
\end{figure*}

\begin{figure*}[!hbt]
\includegraphics[keepaspectratio=true,width=180mm]{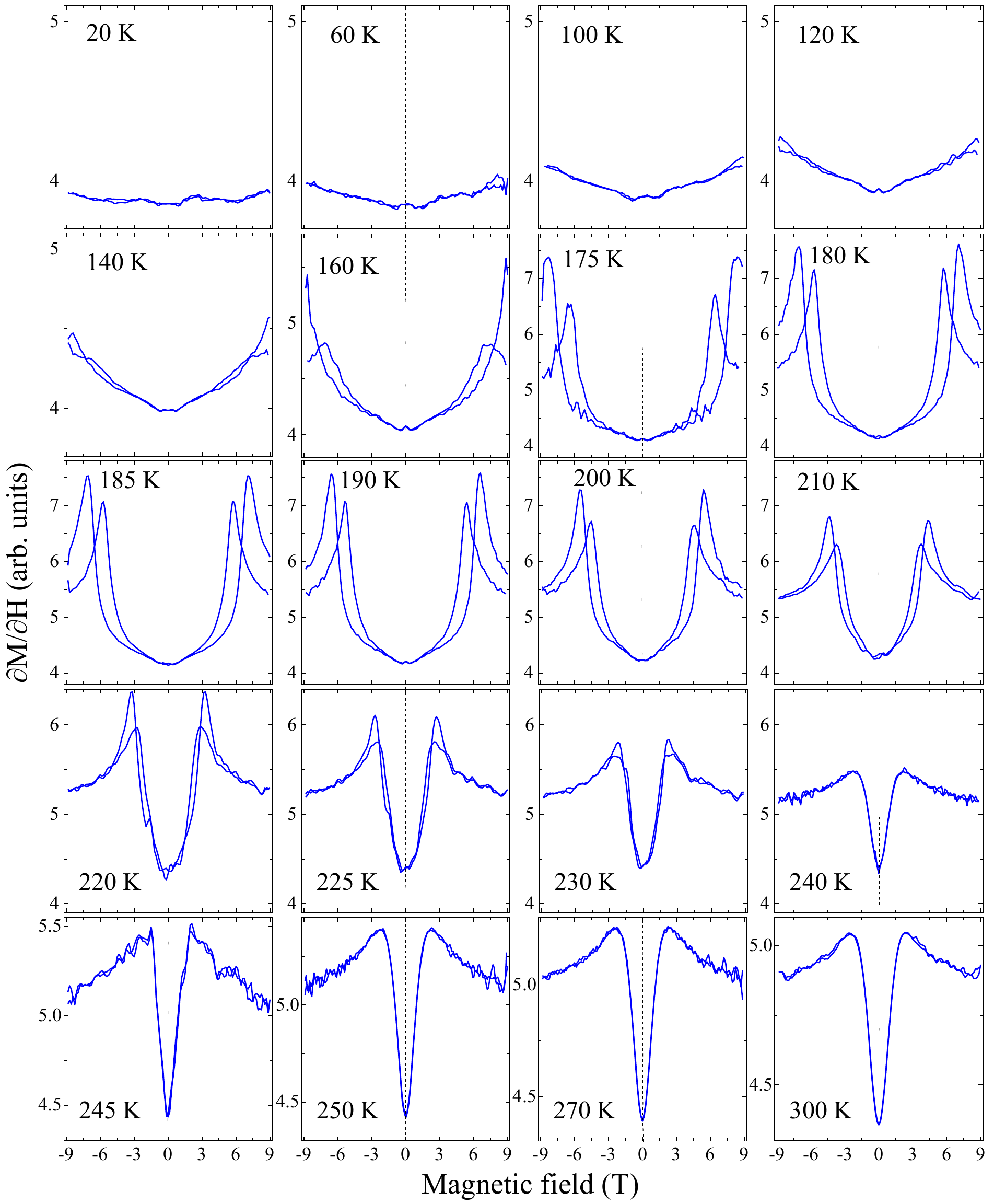}
\vspace{-3mm}
\caption{First derivative curves ${\partial M}/{\partial H}$ of the $M(H)$ loops displayed in Fig. 11.}
\label{Fig5:NPD_evolution}
\end{figure*}

\section{Data and code availability}

The NPD data were generated at the D20 diffractometer of the Institute Laue Langevin (Grenoble, France), and can be accessed at https://doi.org/10.5291/ILL-DATA.5-31-2380. Additional data related to this paper are available from the correspondence authors upon reasonable request. FullProf Suite is available free of charge at www.ill.eu/sites/fullprof.

\bibliography{BIB_YBaCuFeO5}

\end{document}